\documentclass[twocolumn,aps,prl]{revtex4}
\usepackage{graphicx}
\usepackage{ulem}

\begin{document}

\title{Resonance Scattering in Optical Lattices and
Molecules:\\
Interband versus Intraband Effects}
\author{Xiaoling Cui$^{1}$, Yupeng Wang$^{1}$ and  Fei Zhou$^{2}$}
\affiliation{$^{1}$Beijing National Laboratory for Condensed Matter
Physics and Institute of Physics, Chinese Academy of Sciences,
Beijing 100190, China\\
$^{2}$Department of Physics and Astronomy, The University of British
Columbia, Vancouver, B. C., Canada V6T1Z1}
\date{{\small 21 March}}

\begin{abstract}
We study the low-energy two-body scattering in optical lattices with
higher-band effects included in an effective potential, using a
renormalization group approach. The approach captures most
dominating higher band effects as well as all multiple scattering
processes in the lowest band. For an arbitrary negative free space
scattering length($a_s$), a resonance of low energy scattering
occurs as lattice potential depths reaches a critical value
$v_c$; these resonances, with continuously tunable positions $v_c$
and widths $W$, can be mainly driven either by intraband or both
intra- and interband effects depending on the magnitude of $a_s$. We
have also studied scattering amplitudes and formation of molecules
when interband effects are dominating, and discussed an intimate
relation between molecules for negative $a_s$ and repulsively bound
states pioneered by Winkler {\it et al.}\cite{Winkler06}.
\end{abstract}

\maketitle

For a dilute ultracold atomic gas, the two-body s-wave scattering
length $a_s$ is known to be conveniently tunable via
magnetic-field-induced Feshbach resonances\cite{theory,expe}.
Experimentally in the presence of external trapping confinements,
however, binary atomic collision properties can be dramatically
modified as revealed both theoretically and experimentally in
three-dimensional (3D) harmonic
traps\cite{Busch98,Stoferle06,Ospelkaus06}, and in
waveguides\cite{Olshanii,Pricoupenko08,Moritz05,Petrov}. Remarkably,
the waveguide confinement can result in very peculiar
effective potentials as pointed out in a few early
papers\cite{Olshanii,Petrov}; especially, Olshanii {\it et al.}
systematically studied scattering between atoms in a 1D waveguide
and found that the effective potential for atoms in the lowest
transverse mode can reach the hardcore limit.
Interacting atoms in optical lattices are another subject that has attracted
enormous interests for the past few
years\cite{Bloch02,Esslinger05,Ketterle06}. However, till now what
happens to binary collisions in an optical lattice on the other hand
have not been thoroughly studied and the subject of molecules of Bloch waves is also not well
understood. It is becoming essential to understand
the fundamentals of two-body scattering and other few-body physics
of Bloch states; such analyses should form building blocks
for future many-body theories and set potential
references for quantitative calculations of
parameters in many-body Hamiltonians.
Studies of this issue can further cast light on dynamics of colliding atoms or
condensates in optical lattices and coherent control of atoms in
optical lattices\cite{Bloch03}.

Low energy scattering in an optical lattice
was previously investigated and resonance scattering was pointed out
for attractive interactions\cite{Fedichev04}.
Studies there
were carried out in an approximation where laser potentials are approximated as
harmonic ones so that the center-of-mass motion is decoupled from the relative
motion of two scattering atoms; effectively the problem was reduced to two-body
scattering within an individual lattice site, which is
justifiable for deep lattices.
In this Letter to reveal how Bloch waves
are scattered in optical lattices at different depths,
we propose an approach to
resonance scattering without utilizing the approximations of separable
potentials in Ref.\cite{Fedichev04}.
Our approach captures most {\it dominating} higher band effects
as well as all intraband scattering within the lowest band.
It is valid for studies of resonances in deep lattices at small $a_s$
as well as resonances in shallow ones at large $a_s$.
And when applying our approach to lower dimensional
waveguides, we obtain identical results discussed previously
\cite{Olshanii,Petrov}.

Our main proposal is to evaluate $U_{eff}$,
the effective potential for atoms in the lowest band
that takes into account multiple
virtual scattering processes involving {\it higher bands}, and then
apply the same procedure to calculate the full $T$-matrix of
low-energy scattering. When free space scattering lengths $|a_s|$
are comparable to or larger than the lattice constant $a_L$, virtual
scattering to higher bands contributes substantially to scattering
in the lowest band and resonance scattering is driven by both
interband and intraband effects.
We find that the higher-band effects on physical quantities are most pronounced in shallow lattices
near resonances(see Fig.\ref{fig4},\ref{fig5}).
When magnitudes of $a_s$ are arbitrarily small, resonance
scattering at the bottom of the lowest band is predominately driven
by {\it intraband} virtual scattering and is induced mainly by the
enhanced effective masses of atoms in optical lattices.

To facilitate discussions on low-energy scattering atoms, we
start with a two-body Hamiltonian
\begin{eqnarray}
H=\sum_{\alpha} \epsilon_{\alpha} |\alpha\rangle\langle\alpha|
+\sum_{\alpha\beta} U_{\alpha\beta} |\alpha\rangle\langle\beta|,
\end{eqnarray}
with $|\alpha(\beta)\rangle$ being arbitrary two-body scattering
states. For scattering in free space with a short range potential
approximated as $U(\mathbf{r})=U_{0}\delta^3(\mathbf{r})$,
$|\alpha\rangle, |\beta\rangle=|{\bf k},-{\bf k}\rangle$ and
$U_{\alpha\beta}=\frac{U_0}{\Omega}$ with $\Omega$ the volume; to
obtain an effective low-energy Hamiltonian, we employ the
momentum-shell renormalization group(RG) equation
approach\cite{Kaplan98}. The key idea here is at an arbitrary cutoff
momentum $\Lambda$, we can further divide the ${\bf k}$-space into
two regions, i.e., a core region defined by $|{\bf k}| < \Lambda
-\delta \Lambda$ and a shell $\Lambda -\delta \Lambda < |{\bf k}|
<\Lambda$. Correspondingly, we split the Hamiltonian at a given
cutoff $\Lambda$ into three pieces $H(\Lambda)=H^{<}+H^{>}+H^{><}$
which respectively describe interacting atoms within the core,
within the shell and the scattering in between. For atoms with
$|{\bf k}|\ll \Lambda$, the second-order virtual scattering into
high energy states within the shell caused by $H^{><}$ modifies the
low-energy scattering amplitudes and results in a correction
$(\delta U)$ in $H^<$. One can then obtain a differential RG
equation for effective potential $U(\Lambda)$ in $H(\Lambda)$ in
terms of the cutoff $\Lambda$,

\begin{equation}
\frac{1}{U^2}\frac{\delta U}{\delta
\Lambda}=\frac{1}{\Omega}\frac{\delta}{\delta \Lambda}(\sum_{|{\bf
k}|< \Lambda }\frac{1}{2\epsilon_{\bf k}}).\label{RG1}
\end{equation}
So the effective potential $U$ for scattering atoms at momenta
smaller than ${\Lambda}$ is renormalized due to the coupling to
virtual states at larger momenta and is given as
\begin{equation}
\frac{1}{U(\Lambda)}=\frac{1}{U(\Lambda^*)}+\frac{1}{\Omega}\sum_{\Lambda
< |{\bf k}|<\Lambda^*}\frac{1}{2\epsilon_{\bf k}}. \label{Veff}
\end{equation}
Boundary conditions $U(\Lambda^*)=U_0$ and $U(0)=T_0$ relate $U_0$
to the low-energy scattering length $a_s(=\frac{mT_0}{4\pi})$ via
$\frac{m}{4\pi
a_s}=\frac{1}{U_0}+\frac{1}{\Omega}\sum_{|\mathbf{k}|<\Lambda^*}\frac{1}{2\epsilon_{\mathbf{k}}}$.
($\Lambda^*$ is an ultraviolet momentum cut-off that is set by the
range of interactions.)

\begin{figure}[ht]
\includegraphics[width=\columnwidth]{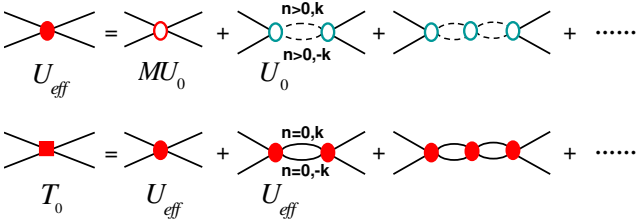}
\caption{(color online). Virtual scattering processes that
contribute to scattering potential $T_0$ in optical
lattices. Different two-particle vertex functions are indicated below
the diagrams. Solid (dashed) lines between two
vertices represent states in the lowest $n=0$ band (higher $n>0$
bands).}\label{fig1}
\end{figure}

One can apply a similar idea to optical lattices. For a cubic
optical lattice with potential
$V(\mathbf{r})=V_0\sum_{i=x,y,z}\sin^2(\frac{2\pi x_i}{\lambda})$
and spacing $a_L=\frac{\lambda}{2}$, Bloch states
$\phi_{n\mathbf{k}}(\mathbf{r})=\frac{1}{\sqrt{\Omega}}\sum_{\mathbf{G}}
w_{n}(\mathbf{k+G})e^{i(\mathbf{k}+\mathbf{G})\cdot\mathbf{r}}$ and
energies $\epsilon_{n\mathbf{k}}$ can be numerically obtained; here
$n$ are band indices; $\mathbf{k}\in BZ$(Brillouin zone) is a quasi-momentum and
$\mathbf{G}$ (and $\mathbf{Q}$ below) are the reciprocal lattice vectors; eigenvector
$w_{n}(\mathbf{q})$ is the Wannier wavefunction in
momentum space. We specify potential depth
$V_0$ in units of the recoil energy $E_R=\frac{\pi^2}{2ma_L^2}$ via
a dimensionless quantity $v=\frac{V_0}{E_R}$.

First we calculate the interaction matrix element between Bloch
states $|\alpha\rangle=|\{m,\mathbf{k}\},\{n,-\mathbf{k}\}\rangle$
and
$|\beta\rangle=|\{m',\mathbf{k'}\},\{n',-\mathbf{k'}\}\rangle$,
\begin{eqnarray}
U_{\alpha\beta}&\equiv&\langle\alpha|U|\beta\rangle=\frac{U_{0}}{\Omega}\sum_{\mathbf{Q}}M_{\alpha}^{\mathbf{Q}\
*}M_{\beta}^{\mathbf{Q}}, \label{matrix}
\end{eqnarray}
here
$M_{\alpha}^{\mathbf{Q}}=\sum_{\mathbf{G}}w_{m}(\mathbf{k+G})w_{n}(\mathbf{Q-k-G})$.
Relevant matrix elements of $U_{\alpha\beta}$ can be classified into
three categories: A) $|\{0,{\bf k}\};\{0,-{\bf k}\}\rangle \leftrightarrow
|\{0,{\bf k}'\};\{0,-{\bf k}'\}\rangle$, i.e. scattering within the lowest band,
$U_{\alpha\beta}$ are given as $MU_0$\cite{M}; these represent the
most dominating processes; B)
$|\{n,{\bf k}\};\{n,-{\bf k}\}\rangle \leftrightarrow |\{n',{\bf k}'\}; \{n',
-{\bf k}'\}\rangle$ with $n\neq0$ or $n'\neq0$ which constitute the most
important scattering processes involving higher bands, give the
next dominating contributions that are approximately equal to $U_0$
(deviations are typically of order of $\frac{v^2}{32}$ or less in shallow lattices); C)
$|\{n, {\bf k}\}; \{n,-{\bf k}\}\rangle \leftrightarrow |\{m',{\bf k}'\};\{n',
-{\bf k}'\}\rangle$ with $m'\neq n'$, i.e. scattering
involving two atoms in different bands; they contribute the
least in shallow lattices (of order of $\frac{v}{8}$ or less)
because of the approximate translational symmetry.

In shallow lattices at an arbitrary $a_s$, we can always neglect
matrix elements in C-class and only keep those in A- and B-class. In
deep lattices near resonances where $a_s$
are small, we keep matrix elements in B-class to remove the
ultraviolet divergence when summing up the virtual scattering to
high energy states; the residue higher band effects after
regularization turn out to be negligible and we again neglect
C-class scattering processes; and our treatments of scattering
processes within the lowest band become exact in this limit.
However, for large $a_s$ and deep lattices that are {\it away} from
the resonances, contributions from C-class scattering
can be comparable to other classes; and by neglecting C-class
contributions, we obtain in this limit estimates only good for
qualitative understanding.
To study resonances, below we adopt a simplest two-coupling-constant model(See Fig.\ref{fig1})
which yields reasonable estimates of higher band effects.

Using the general features of $U_{\alpha\beta}$ discussed above and following the
idea outlined before Eq.(\ref{Veff}), we obtain the effective
potential $U_{eff}$ for the lowest band and further calculate the
scattering potential $T_0$ for states near
$\epsilon_{n\mathbf{k}}=0$, as diagrammatically shown in Fig.1, to
be
\begin{eqnarray}
\frac{1}{U_{eff}}&=&\frac{m \eta}{4\pi
a_LM}(\frac{a_L}{a_s}-C_1),\nonumber\\
\frac{1}{T_0}&=&\frac{m \eta}{4\pi
a_LM}(\frac{a_L}{a_s}-C_1+C_2),\label{T0}
\end{eqnarray}
with $C_{1,2}$ defined as
\begin{eqnarray}
C_1&=&\frac{4\pi
a_L}{m\Omega}\big(\sum_{\mathbf{k}}\frac{1}{2\epsilon_{\mathbf{k}}}-\sum_{n>0,\mathbf{k}}\frac{1}{2\epsilon_{n\mathbf{k}}}\big),\nonumber\\
C_2&=&\frac{4\pi
a_L}{m\eta \Omega}\sum_{n=0,\mathbf{k}}\frac{M}{2\epsilon_{n\mathbf{k}}}.
\label{C2}
\end{eqnarray}
Here $\eta=(1+(1-\frac{1}{M})\frac{U_0}{\Omega}
\sum_{n>0,\mathbf{k}}\frac{1}{2\epsilon_{n\mathbf{k}}})^{-1}$ is
close to unity in the regions that interest us\cite{eta};
evidently $C_1$ and $C_2$ are respectively ascribed to interband
and intraband scattering effects.
Note that when $a_s$ is much bigger than $a_L$, $U_{eff}$ saturates at a value of
$-4\pi a_L M/m C_1$.

\begin{figure}[ht]
\includegraphics[height=5.2cm,width=8cm]{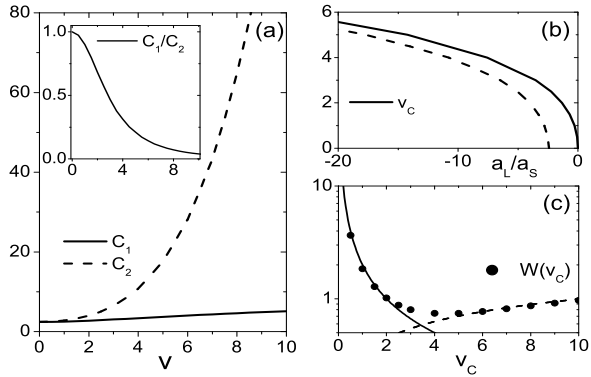}
\caption{(a)Interband($C_1$) and intraband($C_2$) effects vs lattice
potentials $v$. Inset shows the ratio. (b)Resonance position
$v_c$ vs $a_L/a_s$; dashed line is $v_c$ calculated
without interband contributions (i.e. $C_1=0$.) (c)Width $W(v_c)$.
Solid and dashed lines are fit to $\frac{2}{v_c}$
and $\frac{4}{\pi\gamma}\sqrt{v_c}$ respectively.
}
\label{fig2}
\end{figure}

Our results of $C_{1,2}$ are shown in Fig.\ref{fig2}. At $v=0$ and $M=1$, we
reproduce the free space result $T_0=4\pi a_s/m$. With increasing
$v$, the intraband scattering gradually takes a dominating role over
other ones, reflected by a much more rapid increase of $C_2$ than
$C_1$. For instance at $v=5$, $C_1/C_2=0.21$. In the large-$v$
limit, with the lowest band spectrum $\epsilon_{\mathbf{k}}=t\sum_i(1-\cos{k_ia_L})$, $t$ being the hopping
amplitude, we find that
\begin{eqnarray}
C_1=\sqrt{8}v^{\frac{1}{4}},\ C_2= \frac{\pi
\gamma}{32\sqrt{2}}e^{2\sqrt{v}} \ (\gamma\approx 4).
\label{c12_largeV}
\end{eqnarray}

To obtain Bloch wave scattering length $a_{bloch}$ we first
introduce an effective(band) mass
$m_{eff}=1/\frac{\partial^2\epsilon_{nk}}{\partial k ^2}|_0$ and
relate it to the scattering potential $T_0=4\pi a_{bloch}/m_{eff}$.
For a negative $a_s$, a resonance ($a_{bloch}\rightarrow\infty$)
occurs at lattice potential $v_c$ when $\frac{a_L}{a_s}=-(C_2-C_1)$.
Across the resonance, $a_{bloch}$ obeys an asymptotic equation
\begin{eqnarray}
\frac{a_{bloch}}{a_L}=\frac{W(v_c)}{v-v_c}.
\label{esl}
\end{eqnarray}
In the limit
of $|a_s|\ll a_L$,
$v_c$ and $W$ can be estimated using Eq.(\ref{c12_largeV});
in the opposite
limit, they can be obtained
using the perturbation theory with respect to $v$,
\begin{eqnarray}
|a_s|\ll a_L&:&\ \ v_c=\frac{1}{4}\ln^2(\frac{32\sqrt{2}}{\pi
\gamma}\frac{a_L}{|a_s|}), \ W=\frac{4}{\pi\gamma}\sqrt{v_c};\nonumber\\
|a_s|\gg a_L&:&\ \ v_c=2\sqrt{\frac{a_L}{|a_s|}}, \
W=\frac{2}{v_c}.\nonumber
\end{eqnarray}

Both $v_c$ and $W$ are continuously tunable by varying $a_s$. For
ultracold isotopes with negative zero-field scattering lengths such
as $^{85}Rb|2,2\rangle$ $(-390a_0)$, $^{39}K|1,1\rangle$ $(-45a_0)$
and $^{7}Li|2,2\rangle$ $(-27a_0)$, using parameters in
\cite{Bloch09} we find resonances at $v_c=5.5, 10.0, 11.7$
respectively; for $^{87}Rb$ and $^{40}K$ atoms with interspecies
scattering length $a_{bf}=-177a_0$, resonance scattering occurs at
$v_c=7.1$.

For very small $|a_s|$,
$|a_s|\ll \frac{a_L}{C_1}$, the effective potential $U_{eff}$
can be simply related to the
on-site interaction $U_H$ in the Hubbard model,
$\frac{U_{eff}}{\Omega}=\frac{U_H}{N_L}$ ($\Omega=N_La_L^3$).
Following Eq.(5-8), we express $a_{bloch}$ as
\begin{equation}
\frac{a_{bloch}}{a_L}=\frac{1}{4\pi}(\frac{t}{U_H}+\frac{\gamma}{16})^{-1},\label{abloch_TBA}
\end{equation}
which predicts a resonance at $\frac{t}{U_H}\approx-0.25$.

\begin{figure}[ht]
\includegraphics[height=4.8cm,width=8cm]{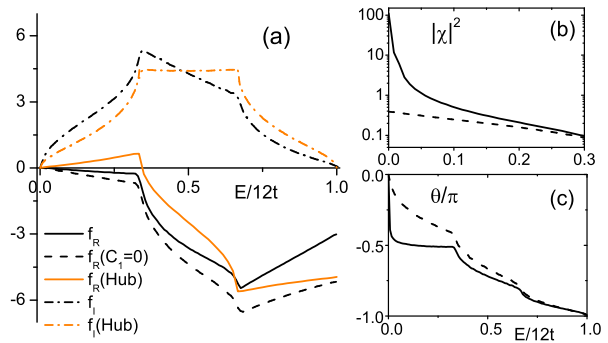}
\caption{(color online). (a) Real($f_R$) and imaginary($f_I$)
components of $f({E})=\frac{4\pi a_L}{m}(T^{-1}(E)-T_0^{-1})$ in an
optical lattice with depth $v=2.5$ (the band width $6t=1.55E_R$).
$f_I$ is related to the normalized density of state
$\rho(\epsilon)$, ${f_I=4 \rho(\frac{E}{2E_R})}$. $f(E)$ calculated
without interband effects ($C_1=0$) or using a Hubbard (Hub) model
are also shown. (b,c) Scattering cross sections($4\pi a^2_L
|\chi|^2$, near the bottom of the band) and phase shifts($\theta$)
at $a_s=-0.5a_L$ (resonance position $v_c=2.37$) and $v=2.5$,
calculated using T-matrix,
$|\chi|e^{i\theta}=\frac{m_{eff}T(E)}{4\pi a_L}$. Dashed lines are
the data without interband effects.}\label{fig4}
\end{figure}

We now turn to scattering matrix $T(E)$ for two atoms with total
energy $E$\cite{Messiah}. Using identical diagrams as shown in
Fig.1, one can introduce a Lippmann-Schwinger equation for two
scattering atoms in optical lattices; the solutions for $T(E)$ can
be obtained as
\begin{eqnarray}
\frac{1}{{T(E)}}&=&\frac{m \eta(E)}{4\pi
a_LM}(\frac{a_L}{a_s}-C_1(E)+C_2(E)),\label{TE}
\end{eqnarray}
where $C_{1,2}(E)$ can be obtained by substituting
$\epsilon_{n\mathbf{k}}$ in $C_{1,2}$ of Eq.(\ref{T0}),(\ref{C2})
with $\epsilon_{n\mathbf{k}}-E/2-i0^+$. At small $E$, the first two
terms in the bracket in Eq.(\ref{TE}) can be approximated to be
$U_{eff}$ that dictates the low energy scattering. Fig.\ref{fig4}
shows cross sections and phase shifts in shallow lattices where
higher band effects are dominating (See (b),(c)). When $E$ is
approaching zero, asymptotically we have
$T^{-1}(E)=\frac{m_{eff}}{4\pi}(a^{-1}_{bloch}+\beta k^2_E a_L+
ik_E),\ k_E= \sqrt{m_{_{eff}}E}$ and $\beta$ approaches $0.03\pi$
when $C_1$ is negligible. T-matrix and scattering phase shifts in
optical lattices exhibit much richer $E$-dependence than in free
space; this is mainly due to a relatively large range of effective
interactions (of order of $a_L$) in the lowest band compared to that
of free space resonances, or a small resonance energy width (of
order of $t$ as suggested in Eq.(\ref{abloch_TBA})).

\begin{figure}[ht]
\includegraphics[height=4.2cm,width=8.5cm]{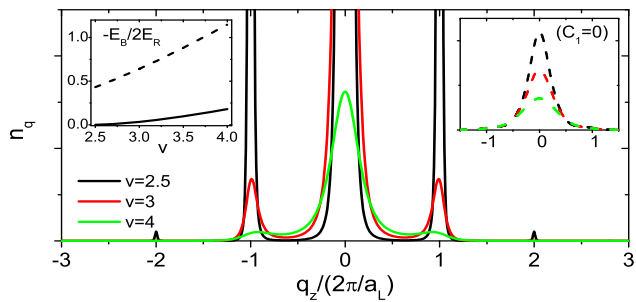}
\caption{(color online). Real momentum distribution
$n_{\mathbf{q}}$($q_x=q_y=0$, normalized) of bound states
at different potential depths $v(>v_c)$; $a_s=-0.5a_L$ and the resonance occurs at
$v_c=2.37$.
Right inset shows results without higher band effects(i.e. $C_1=0$) and
the left one shows the binding energy (dashed line is the estimate without $C_1$);
note that neglecting higher band effects severely overestimates $|E_B|$ leading to a
much less singular momentum distribution function $n_{\mathbf{q}}$.
}\label{fig5}
\end{figure}

Beyond $v_c$, a stable molecule can be formed with a binding energy
$E_B(<0)$. $E_B$ can be obtained by solving the following two-body
equation
\begin{eqnarray}
0&=&\frac{m \eta}{4\pi
a_LM}(\frac{a_L}{a_s}-C_1(E_B)+C_2(E_B)),\label{EB}
\end{eqnarray}
Near resonances, $|E_B|$ is proportional to $a_{bloch}^{-2}$ (see
Eq.(\ref{esl}) for $a_{bloch}$) with the same scaling dimension as in free space. For
a bound state
$|\Psi\rangle=\sum_{n\mathbf{k}}c_{n\mathbf{k}}\psi^{\dag}_{n\mathbf{k}}\psi^{\dag}_{\mathbf{n-k}}$
, $n_{\mathbf{q}}=\sum_{n \mathbf{k G}} \delta_{\mathbf{q,k+G}}\
|w_n(\mathbf{k+G})|^2 |c_{n\mathbf{k}}|^2$ and $c_{n\mathbf{k}}$ is
proportional to $1/(2\epsilon_{n\mathbf{k}}-E_B)$. In Fig.\ref{fig5},
we plot $n_{\mathbf{q}}$ for $a_s=-0.5a_L$ where higher band effects
are dominating.

In the limit of deep lattices\cite{Winkler06,Wouters06},
one can neglect higher band effects by setting $C_1(E)=0$ in
Eq.(\ref{TE}),(\ref{EB}) and $\eta(E)=1$.
Both the T-matrix and the
binding energy in this limit exhibit a {\it generalized}
particle-hole symmetry due to a property of the single particle
density of states, $\rho({\epsilon})=\rho(6t-\epsilon)$. So for a
given scattering length $a_s$, one finds that $Re T^{-1}(E)+Re
T^{-1}(12t-E)=Re T^{-1}(12t)+Re T^{-1}(0)$ and $Im
T^{-1}(E)=Im T^{-1} (12t-E)$(see Fig.\ref{fig4}a). More
important, the stable molecules below the lowest band for negative
scattering lengths $a_s(<0)$ have close connections to mid-gap
repulsively bound states for positive scattering lengths
that were first thoughtfully pointed out by Winkler {\it et al.}\cite{Winkler06}. In
addition, the T-matrix for negative $a_s$ can also be related to
that for positive $a_s$ via a simple reflection symmetry. Indeed, by
examining Eq.(\ref{TE}), (\ref{EB}) in the limit of deep lattices we
verify the following exact relations between $a_s(<0)$ and $-a_s(>0)$ cases,
$T(E,a_s)= -T^*(12t-E, -a_s)$, $-E_B(a_s)=E_B(-a_s)-12t$;
resonance scattering and bound states near the bottom of lowest band
for a negative $a_s$ therefore imply resonance scattering and
bound states near the the top of the band for a positive scattering
length $-a_s$. Note that our equation for the repulsively bound states in
this particular limit is identical to the one in
Ref.\cite{Winkler06}.

In conclusion, we have developed an approach to low-energy resonance scattering
in optical lattices taking into account not only
the intraband physics but more importantly
higher band effects. The resonance scattering in optical lattices offers an
alternative path to unitary cold Bose gases so far mainly studied
via Feshbach resonances\cite{Jin08}. Resonances can also be
utilized to study exciting few-body physics of heteronuclear
molecules\cite{Petrov07} and Efimov states. We thank Immanuel Bloch, Hanspeter B{\"
u}chler, Gora Shlyapnikov, Victor Gurarie, Maxim Olshanii, Dmitry
Petrov, Leo Radzihovsky and Ruquan Wang  for stimulating discussions
and the KITPc 2009 cold atom workshop in Beijing for its
hospitality. This work is in part supported by NSFC, $973$-Project
(China), and by NSERC (Canada), Canadian Institute for Advanced
Research.


\begin{references}{}
\bibitem{theory}E. Tiesinga {\it et al.}, Phys. Rev. A {\bf 47}, 4114
(1993); J. P. Burke {\it et al.}, Phys. Rev. Lett. {\bf 81}, 3355
(1998).

\bibitem{expe}S. Inouye {\it et al.}, Nature {\bf 392}, 151 (1998);
Ph. Courteille {\it et al.}, Phys.
Rev. Lett. {\bf 81}, 69 (1998); J. L. Roberts {\it et al.}, Phys.
Rev. Lett. {\bf 81}, 5109 (1998).


\bibitem{Busch98}T. Busch {\it et al.}, Found. Phys. {\bf 28}, 549 (1998).

\bibitem{Stoferle06}T. St{\"o}ferle {\it et al.}, Phys. Rev. Lett. {\bf 96}, 030401 (2006).
\bibitem{Ospelkaus06}C. Ospelkaus {\it et al.}, Phys. Rev. Lett. {\bf 97}, 120402 (2006).

\bibitem{Olshanii}M. Olshanii, Phys. Rev. Lett. {\bf 81}, 938
(1998); T. Bergeman {\it et al.}, Phys. Rev. Lett. {\bf 91}, 163201
(2003).
\bibitem{Petrov}D. S. Petrov {\it et al.}, Phys. Rev. Lett.
{\bf 84}, 2551 (2000); D. S. Petrov {\it et al.}, Phys. Rev. A {\bf
64}, 012706 (2001).
\bibitem{Pricoupenko08}L. Pricoupenko, Phys. Rev. Lett. {\bf 100}, 170404 (2008).
\bibitem{Moritz05}H. Moritz {\it et al.}, Phys. Rev. Lett. {\bf 94}, 210401 (2005). 





\bibitem{Bloch02}M. Greiner {\it et al.}, Nature {\bf 415}, 39 (2002).
\bibitem{Esslinger05}M. K{\"o}hl {\it et al.}, Phys. Rev. Lett. {\bf 94}, 080403 (2005);
R.B. Diener and T. L. Ho, Phys. Rev. Lett. {\bf 96}, 010402(2006).


\bibitem{Ketterle06}J. K. Chin {\it et al.}, Nature {\bf 443}, 961 (2006).

\bibitem{Bloch03}O. Mandel {\it et al.}, Nature {\bf 425}, 937 (2003).

\bibitem{Fedichev04}P. O. Fedichev {\it et al.}, Phys. Rev. Lett. {\bf 92}, 080401 (2004).

\bibitem{Kaplan98}
A systematic approach was originally
proposed in D. B. Kaplan {\it et al.}, Nucl. Phys. B {\bf 534}, 329(1998).



\bibitem{M}Numerical results show that the lowest band interaction $U_{\alpha\beta}$
have relatively weak dependence on $\alpha,\beta$; coefficient $M$
calculated for $\mathbf{k=k'=0}$ follows $1+\frac{3v^2}{32}$ and
$(\frac{\pi}{2})^{\frac{3}{2}}v^{\frac{3}{4}}$ respectively in the
small and large $v$ limit.

\bibitem{eta}
$\eta=1$ in free space, but depends on $a_s$ in an optical
lattice. Deviation $|\eta-1|/\eta$ are negligible near
resonances, i.e., large $|a_s|$ for shallow $v$ or small
$|a_s|$ for deep $v$.


\bibitem{Bloch09}Th. Best {\it et al.}, Phys. Rev. Lett. {\bf 102}, 030408 (2009).

\bibitem{Messiah}A. Messiah, {\it Quantum Mechanics} (Dover Publications, 1999).






\bibitem{Winkler06}K. Winkler {\it et al}, Nature {\bf 441}, 853(2006).
We are thankful to I. Bloch, H. B{\"u}chler and P. Zoller for
drawing our attention to repulsively bound states.

\bibitem{Wouters06}
This limit was also studied in M. Wouters {\it et al.},
Phys. Rev. A {\bf 73}, 012707(2006).
For 1D exact numerical results, see
G. Orso {\it et al.}, Phys. Rev. Lett. {\bf 95}, 060402 (2005).



\bibitem{Jin08}S. B. Papp {\it et al.}, Phys. Rev. Lett. {\bf 101},
135301 (2008); S. E. Pollack {\it et al.}, Phys. Rev. Lett. {\bf 102}, 090402 (2009).

\bibitem{Petrov07}D. S. Petrov {\it et al.}, Phys. Rev. Lett. {\bf 99}, 130407 (2007).

\end{references}
\end{document}